\pdfoutput=1
\documentclass[11pt]{article}
\usepackage[utf8]{inputenc}
\usepackage{times}

\usepackage[margin=1in]{geometry}
\usepackage{microtype}
\usepackage{amsmath,amsthm,amssymb}
\usepackage{graphicx}
\usepackage{booktabs}
\usepackage{array}
\usepackage{tabularx}
\usepackage{xcolor}
\usepackage{caption}
\usepackage{flafter}
\usepackage{float}
\usepackage{placeins}

\PassOptionsToPackage{hyphens}{url}
\usepackage[colorlinks=true,linkcolor=black,citecolor=blue!50!black,urlcolor=blue!50!black]{hyperref}

\hypersetup{
  pdftitle={Separating Measurement Failures from Defence Failures in Split Language-Model Privacy},
  pdfauthor={Georgios Politis and Evangelos Pappas},
  pdfsubject={Retrospective evidentiary reanalysis of split language-model privacy evaluations}
}
\newtheorem{proposition}{Proposition}
\expandafter\newcommand\csname artifactnameE1\endcsname{decoy-gradient records}
\expandafter\newcommand\csname artifactnameB1\endcsname{boundary-review records}
\expandafter\newcommand\csname artifactnameR1\endcsname{exploration records}
\expandafter\newcommand\csname artifactnameP4\endcsname{pilot records}
\expandafter\newcommand\csname artifactnameE8\endcsname{wrapper records}
\expandafter\newcommand\csname artifactnameER\endcsname{deployment records}
\expandafter\newcommand\csname artifactnameMI\endcsname{MI-analysis records}
\DeclareRobustCommand{\artifact}[1]{\hyperlink{artifact-#1}{\csname artifactname#1\endcsname}}
\newcommand{\pp}{\,\mathrm{pp}}
\newcommand{\repourl}{\url{https://github.com/setloop-io/Separating-Measurement-Failures-from-Defence-Failures-in-Split-Language-Model-Privacy}}
\newcommand{\RoneCells}{19}
\newcommand{\RoneQualifiedBreaches}{13}
\newcommand{\RoneUtilityFailures}{5}
\newcommand{\RoneGainFailures}{7}
\newcommand{\RoneUndetermined}{6}
\newcommand{\PfourCells}{7}
\newcommand{\PfourUtilityFailures}{2}
\newcommand{\PfourPrivacyFailures}{4}
\newcommand{\PfourPrivacyUndetermined}{3}
\newcommand{\PinnedInputs}{122}
\newcommand{\ConsistencyChecks}{1,522}

\title{\textbf{Separating Measurement Failures from Defence Failures in Split Language-Model Privacy}}
\author{Georgios Politis\\ Setloop.io\\ {\small\texttt{geo.politis@gmail.com}} \and
Evangelos Pappas\\ Setloop.io\\ {\small\texttt{evanz.pappas@gmail.com}}}
\date{}
\begin{document}
\maketitle
\begin{abstract}
Split-training privacy evaluations fail through distinct mechanisms.
Evaluators can overlook an exposed gradient channel, align reconstructions
against shifted labels, or deploy uncalibrated probes before testing a
candidate defence, while calibrated attacks can also refute real defences.
Separating these failure modes alters which claims survive. We conduct a
retrospective evidentiary reanalysis across four split-LLM campaigns and
two cross-cutting scope checks covering host key custody and variational
mutual-information bounds, mapping every recorded claim to one of five
admissibility states (breach, non-detection, undetermined coverage, invalid
measurement, or out-of-scope inference). Across a \RoneCells-configuration
exploration, only four of 24 required probes qualified, yet those four
calibrated probes still established qualified privacy breaches in
\RoneQualifiedBreaches\ configurations, while utility or learning-gain
failures eliminated the remaining \RoneUndetermined. In a seven-cell
defence-redesign pilot, \PfourUtilityFailures\ quantised configurations
failed utility gates, calibrated next-byte structure probes breached the
$0.02$ gate on all \PfourCells\ cells ($0.097$ to $0.112$), and qualified
punctuation-property probes reached $0.79$ to $0.89$, yielding
\PfourPrivacyFailures\ failed and \PfourPrivacyUndetermined\ undetermined
composite privacy assessments (the latter from marginal null-band
excursions on punctuation bindings). Negative evidence against both defence
families remains intact. Scope checks further show that shared-key
orthogonal wrappers protect passive network captures without hiding
activations from a compromised restoring host, and why near-zero
variational lower bounds leave worst-case leakage unconstrained. Rerunnable
scripts verify \ConsistencyChecks\ checks across \PinnedInputs\
SHA-256-pinned artifacts while preserving each campaign's native seed
intervals and retrospective boundaries.
\end{abstract}

\section{Introduction}
\label{sec:intro}

Split learning partitions a neural network between a local client and a
remote server so raw training text stays on the client device
\cite{vepakomma2018,splitsurvey2026}. Intermediate activations and returned
loss gradients still cross that boundary. Gradient and representation
inversion attacks recover private tokens from both exchanged tensors
\cite{gisurvey2022,dager2024,fedspy2026}, and deeper cut points retain
recoverable internal states \cite{depth2025,actinv2026,bettisplit2026}.
Interpreting any reported recovery score requires exact experimental
context. Attack outcomes shift substantially with the messages an adversary
observes, the pretrained or fine-tuned weights and cryptographic keys it
possesses, the token index it targets, and the calibration controls backing
the evaluator \cite{ablation2026,adaptive2022,labelbudget2026}.

Recent empirical studies expose individual pieces of this measurement
problem. Omitting a client-returned gradient channel invalidates privacy
assurance derived from forward activations alone \cite{paper1}, while adding
backward gradients shifts exact-document recovery far more sharply than
average token accuracy \cite{gradients2026}. Choice of aggregation unit
alters the substantive conclusion. Although LLM-Attack Suite standardises
observation contracts, undefended controls, and saved predictions
\cite{suite2026}, and Smith et al.\ \cite{mia2026} show how adaptive
optimisation and evaluator choices reshape inversion tradeoffs,
multi-campaign studies still require explicit rules reconciling conflicting
claims. Empirical privacy auditing in differential privacy similarly
separates valid lower bounds on leakage from unproven upper assurances
\cite{jagielski2020}.

This paper re-examines four split-LLM training campaigns and two
cross-cutting security claims drawn from one research programme. Individual
campaigns investigate decoy-padded gradients, shifted token labels in a
boundary-replacement study, incomplete probe qualification across a
\RoneCells-cell exploration, and a calibrated seven-cell defence-redesign
pilot. Each campaign failed for a distinct reason. Two companion scope
analyses then test whether shared-key orthogonal wrappers withstand a
compromised restoring host \cite{permattack2025} and whether near-zero
variational mutual-information readouts bound worst-case leakage
\cite{belghazi2018,mcallester2020,song2020}. Shared corpora and sequential
design choices require keeping each campaign's native estimand and seed
unit separate.

Our reanalysis traces every headline statement from the adversary's
observation surface back to the primary generated tables. Five
admissibility outcomes separate confirmed breaches from non-detection,
undetermined coverage, invalid targets, and out-of-scope threat transfers.
Distinguishing broken probes from broken defences matters. Executable
scripts reconstruct 76 configuration-by-seed contrasts from the exploration
alongside paired losses and candidate probe advantages from 66
defence-pilot runs and verify \ConsistencyChecks\ internal checks across
\PinnedInputs\ SHA-256-pinned artifacts \cite{artifacts2026}. Archived
session-identity logs, throughput timings, and key-derivation manifests
bound what shared-key rotation achieves in practice, while an analytic
lower-bound argument delimits variational mutual-information readouts.

\section{Scope and retrospective admissibility framework}
\label{sec:design}

A stated privacy claim paired with its supporting experimental records forms
our basic unit of analysis. Scope covers four training campaigns and two
security-transfer arguments. Because we selected these episodes from an
earlier programme synthesis whose outcomes were already known, the resulting
archive traces how local measurements grew into sweeping claims while leaving
field-wide error rates unestimated. Component results remain credited to
their originating studies, and Table~\ref{tab:master-audit} summarises how
each recorded headline maps onto its admitted evidentiary state.

All historical runs finished before we specified this reconciliation.
Throughout the text, frozen exploratory protocols, completed pilot
executions, unexecuted confirmation proposals, and secondary consistency
checks remain explicitly separated. Our analysis recomputes stored loss
contrasts, probe advantages, and decision thresholds directly from archived
tables without retraining victim models or fitting stronger attackers.
Prospective validation requires an independent cohort. Where LLM-Attack
Suite defines observation contracts and control conventions
\cite{suite2026}, we apply its reporting structure to compare cross-campaign
claims under their original experimental budgets.

\paragraph{Evidence required for a claim.}
Evaluating any candidate claim begins by logging the split protocol, victim
cohort, secret-key custody, remote host state, and observable tensor
channels. Target indexing and control bindings come next. Specifically, each
record pairs the protected token or property target with its exact tensor
index, training horizon, baseline guesser, acceptance gate, and independent
resampling unit. A qualification binding ties positive canary floors and
permutation-null bands to one concrete target, channel, defence family, and
probe implementation. Passing controls on one channel leaves unmeasured
channels unqualified.

Primary generated CSV and JSON outputs alongside active collection manifests
supersede informal narrative summaries. Whenever a subsequent technical
review revises an earlier disposition, we report both the updated status and
every attached operational caveat regardless of which reading looks more
favourable. Every retained number carries explicit provenance.
Appendix~\ref{app:artifacts} binds each quantity to its SHA-256-hashed file
path and marks whether our scripts recomputed the statistic from raw seed
records or extracted it from a cited study.

\paragraph{Five admissibility outcomes for a specified claim.}
Individual endpoints within a single split configuration often merit
distinct evidentiary labels.
\begin{description}
  \item[Breach.] An admissible, calibrated probe exceeds its frozen
    acceptance threshold and establishes leakage for the specified observer,
    target, and finite protocol. Unprobed disclosures may still coexist
    alongside it.
  \item[Non-detection.] A calibrated attack stays within its acceptance rule
    under the tested architecture, dataset, and optimisation budget.
    Consistent with empirical differential-privacy auditing
    \cite{jagielski2020}, staying below an attack threshold leaves
    worst-case mutual information and adaptive recovery unbounded from above.
  \item[Undetermined.] Missing coverage or failed controls block privacy
    passes.
  \item[Invalid.] A verified implementation defect severs the recorded score
    from the claimed estimand, as occurs when next-token predictions are
    misread as current-token reconstructions. Properly relabelled, the
    underlying number may remain informative.
  \item[Out-of-scope.] Transferring a measurement to a stronger observer,
    altered key-custody regime, or untested protocol exceeds what the
    original experiment can support.
\end{description}

Utility loss and causal remote-learning gain are tracked alongside privacy
under each campaign's original conjunctive decision rule. Conjunctive gating
acts asymmetrically across endpoints. Within the \RoneCells-cell
exploration, one qualified breach or one utility failure suffices to reject
a candidate even when companion probes remain undetermined, whereas
incomplete qualification alone blocks any joint privacy pass. Failing an
exploratory remote-gain threshold similarly leaves population learning
benefit unconfirmed on that split.

\paragraph{Transferring a conclusion.}
Carrying an empirical privacy result into a new deployment setting requires
matching protocols, secret-key custody, observed tensors, target alignments,
and calibrated probe bindings. Wire confidentiality differs sharply from
host confidentiality. When an authorised server holds a shared orthogonal
key to restore activations in memory, passive wire non-detection says
nothing about a compromised host reading its own RAM \cite{permattack2025}.
Correlated token positions within fixed evaluation panels likewise add zero
independent training seeds and restrict how far finite exploratory
intervals generalise.

\begin{table}[t]
\centering\footnotesize
\setlength{\tabcolsep}{4pt}
\begin{tabularx}{\linewidth}{@{}>{\raggedright\arraybackslash}p{0.17\linewidth}>{\raggedright\arraybackslash}p{0.19\linewidth}>{\raggedright\arraybackslash}p{0.19\linewidth}>{\raggedright\arraybackslash}p{0.17\linewidth}>{\raggedright\arraybackslash}X@{}}
\toprule
Investigation & Evaluated view & Initial claim & Admitted state & Retained quantitative basis \\
\midrule
Decoy gradients (\artifact{E1}) & Forward activations vs.\ joint gradient view & Passing forward check implies boundary privacy & Out-of-scope and Breach & Exact real/decoy row recovery on 4,096/4,096 frames per run (9 seeds), $+0.65$ to $+1.50\pp$ token gain, $0.01$-nat clip/noise utility cost \\
\addlinespace[3pt]
Boundary review (\artifact{B1}) & Representations $(x_1,\ldots,x_{T-1})$ vs.\ $(x_2,\ldots,x_T)$ & Trained boundary resists input reconstruction & Invalid for current token & Shifted label scores next-token prediction, 6/6 zero-return controls beat remote, answer match drops $34.79\%$ to $18.15\%$ \\
\addlinespace[3pt]
19-cell exploration (\artifact{R1}) & 24 required probes (3 targets, 4 channels, 2 families) across 4 seeds & 19 rejections stem from 20/24 uncalibrated probes & Breach (\RoneQualifiedBreaches) and Undetermined (\RoneUndetermined) & 4/24 probes qualify, \RoneQualifiedBreaches\ breaches, \RoneUtilityFailures\ utility and \RoneGainFailures\ gain failures, \texttt{span-6} passes utility (mean delta $0.099971$ nats) \\
\addlinespace[3pt]
Defence pilot (\artifact{P4}) & Punctuation-property and next-byte structure probes across 3 paired seeds & $40\times$ excess implies broken gate or probe & Breach on all \PfourCells; composite failed (\PfourPrivacyFailures) / undetermined (\PfourPrivacyUndetermined) & \PfourUtilityFailures/\PfourCells\ fail utility, 7/7 pass gain; qualified punctuation $0.79$--$0.89$ and next-byte $0.097$--$0.112$ vs.\ $0.02$ gate \\
\addlinespace[3pt]
Wrapper and ratchet (\artifact{E8}, \artifact{ER}) & Passive wire capture under shared cloud key $Q$ & Zero-cost protection against compromised cloud & Non-detection (ER wire; low E8 wire score) and Out-of-scope (host) & $3.71\%$ pooled depth-1 wire score ($5.73\%$ max session), cloud holds $Q$, \texttt{N=16} ratchet cuts throughput by $63.16\%$ (20/24 identity) \\
\addlinespace[3pt]
Variational MI (\artifact{MI}) & Donsker--Varadhan critic readout $L_T$ on defended boundary & Near-zero MINE certifies small leak budget & Out-of-scope for upper ceiling & $L_T \le \operatorname{MI}(X,Z)$ bounds leakage from below only; $T=0$ yields $L_T=0$ when $Z=X$ leaks 8 bits ($m=256$) \\
\bottomrule
\end{tabularx}
\caption{Cross-campaign admissibility summary across all six re-examined
investigations. Each row pairs the evaluated observation surface and initial
claim with the admitted evidentiary state from Section~\ref{sec:design} and
its retained quantitative basis (pp denotes percentage points; MINE denotes
Mutual Information Neural Estimation \cite{belghazi2018}). Hyperlinked record
names point to the SHA-256-hashed provenance entries in
Appendix~\ref{app:artifacts}.}
\label{tab:master-audit}
\end{table}

\section{Four campaign case studies}
\label{sec:cases}

\subsection{The decoy-gradient experiment}
\label{sec:f1}

Padding batches with synthetic decoy rows aims to obscure which activation
vectors correspond to real user tokens \cite{paper1}. During each split
step, the trusted client transmits protected forward activations to the
cloud, receives remote hidden states, evaluates the private training loss
locally, and sends output gradients back to the cloud (\artifact{E1}).
Decoy rows receive identically zero gradient. Because the local loss
ignores synthetic positions, that deterministic zero-support mask exposes
the exact real-versus-decoy row partition without revealing token
identities directly, much as gradient-sparsity patterns leak metadata in
federated vision and trajectory settings \cite{gisurvey2022,trace2026}.

Archived logs document exact partition recovery across all 4,096 inspected
frames \emph{per run} over nine training seeds (\artifact{E1}). Recovering
token identities requires a separate content attack. On that distinct token
denominator, a frequent-word probe beats a constant-guess baseline by
$+0.65$ to $+1.50\pp$. Across six utility-compliant cells that expose
backward gradients (three seeds on each of two datasets),
forward-activation probes stay inside the $+1.0\pp$ privacy gate fixed
prior to execution, whereas joint forward-and-backward evaluators breach
that gate decisively.

Forward-only non-detection remains accurate for an observer restricted to
forward activations. Extending that assurance to the cloud's actual
bidirectional view constitutes an out-of-scope inference that the recorded
backward breach refutes. Clipping each gradient row and adding Gaussian
noise eliminates the scored partition leak across all 1,024 inspected
frames per run over three mitigation seeds (producing $0.400$ partition
agreement at a $32/80$ base rate) while imposing roughly $0.01$ nats of
held-out cross-entropy \emph{utility cost} (\artifact{E1}). That
$0.01$-nat figure measures validation loss degradation on the held-out
split, while multi-step gradient accumulation and exact
batch-reconstruction attacks \cite{dager2024,fedspy2026} remain untested
against the noisy channel. Even within that bounded probe scope, closing a
deterministic support leak via per-row clipping and noise refutes blanket
claims that perturbation mechanisms never mitigate training-time leakage.

\subsection{The boundary-replacement review}
\label{sec:f2}

Reviewing the boundary-replacement codebase uncovered an off-by-one
indexing shift between captured representations and target token IDs
(\artifact{B1}). Hidden states for $(x_1,\ldots,x_{T-1})$ met labels
$(x_2,\ldots,x_T)$. Every activation row encoding current token $x_t$ was
consequently evaluated on how accurately a probe predicted the subsequent
token $x_{t+1}$. High canary scores under that shifted alignment validate
next-token predictive signal alone and invalidate claims of resistance to
current-token input reconstruction, though the raw numbers remain
interpretable if explicitly relabelled as next-token probes.

Independent evidence-review records from the same study also re-examined
whether remote computation improved model quality (\artifact{B1}).
Source-reported comparisons across six clean corpus-and-seed pairs show
that standalone client baselines trained from scratch with zero remote
return outperform the split-trained remote model every time. Learned remote
layers were consequently unnecessary for those evaluated endpoints. Zeroing
the return pathway of an already trained split checkpoint tests a different
counterfactual, since co-adapted client layers naturally degrade when their
expected remote input disappears mid-inference.

Checkpoint-level scoring exposed an equally sharp divergence between
continuous loss and discrete generation metrics (\artifact{B1}). Continuous
loss cleared its threshold. Held-out negative log-likelihood rose by roughly
$0.034$ nats within rule, yet teacher-forced top-1 match on answer tokens
dropped $16.64\pp$ from $34.79\%$ to $18.15\%$. Answer-token matching under
teacher forcing differs from standard multi-choice benchmark accuracy and
comes here from a single reviewed checkpoint. Paired with the unmeasured
current-token reconstruction risk, this discrepancy narrows both the
delegation and utility conclusions supported by the archive.

\subsection{The privacy, utility and gain exploration}
\label{sec:f3}

Across 232 training runs spanning \RoneCells\ candidate configurations and
four seeds per cell, the exploratory trilemma campaign rejected every
configuration under its frozen rules (\artifact{R1}). Recomputing the cell
summary yields \RoneQualifiedBreaches\ qualified privacy breaches,
\RoneUtilityFailures\ utility failures, and \RoneGainFailures\ remote-gain
failures. Rejection causes overlap across cells. All \RoneUndetermined\
configurations whose privacy status remained undetermined failed utility or
learning-gain gates independently, including \texttt{width-64} on gain
alone.

Only four of the 24 required privacy probes passed pre-test qualification.
Those four calibrated instruments consisted of input-byte probes operating
on inbound and outbound gradients across the two evaluated defence families
(\artifact{R1}). Unqualified probes prevent certifying a joint privacy
pass, yet breaches triggered by the four calibrated probes remain
scientifically valid alongside the independent utility and gain failures.

Boundary proximity on the \texttt{span-6} cell illustrates why reanalysis
scripts must check exact floating-point inequalities (\artifact{R1}).
Across four training seeds, \texttt{span-6} recorded a mean utility delta
of $0.0999705344$ nats ($2.95\times 10^{-5}$ nats inside the $0.10$ limit)
and an upper 95\% Student-$t$ bound of $0.1195024415$ nats (inside the
$0.12$ cap). Both utility gates passed. Saved rejection flags instead trace
to learning-gain failure paired with a qualified byte-probe breach across a
four-seed exploratory cohort unadjusted for multiplicity.

An accompanying confirmation proposal specified eight seeds per candidate
(\artifact{R1}). Programme leads never froze, authorised, or ran that
stage. Using plug-in exploratory means and variances, a prospective
calculation bounded whole-family confirmation power at $0.263772$ (dropping
sharply under wider variance multipliers) and served strictly as a planning
stop rule for future compute allocation.

\subsection{The defence-redesign pilot}
\label{sec:f4}

Rebuilt probes drove a subsequent seven-cell defence-redesign pilot
comprising 66 runs across three paired seeds per cell and evaluation panels
of 16,384 token windows (\artifact{P4}). Zero candidates achieved pilot
eligibility. Every cell cleared the causal remote-gain gate and five
cleared utility, whereas one-bit and two-bit activation quantisation failed
their utility limits. Under the campaign's all-bindings aggregation policy,
composite privacy assessments recorded \PfourPrivacyFailures\ failures and
\PfourPrivacyUndetermined\ undetermined cells.

Calibrated atomic probes nonetheless captured severe leakage across the
tested quantisation and noise mechanisms (\artifact{P4}). Across qualified
seed bindings, worst-seed sequence-property (punctuation-fraction)
correlation advantages spanned $0.79$ to $0.89$ against an acceptance
ceiling of $0.02$ (roughly $40\times$ the gate), while next-byte structure
(five-class accuracy) advantages reached $0.097$ to $0.112$ across all
seven cells (roughly $5\times$ the gate, with all 21 of 21 trained seed
bindings qualified). Unlike Section~\ref{sec:f2}, next-byte class served
here as an intentional target. Observing calibrated probe advantages well
above a policy threshold refutes the candidate defences under that rule
without implying probe malfunction. Recent split-LLM obfuscation studies
report comparable persistence under perturbation-only defences
\cite{privpair2026,mixguard2026}.

Calibration source files separate four numerical constants by role
(\artifact{P4}). Positive-control canary floors sit at $0.10$ for sequence
properties (punctuation fraction) and $0.02$ for next-byte structure
targets, whereas both the permutation-null band and the candidate
acceptance gate equal $0.02$; semantic targets beyond byte classes were
explicitly omitted from the pilot protocol. Passing a canary floor confirms
probe responsiveness on unprotected controls, and qualification logs record
the four marginal null-band excursions on punctuation-property bindings
(across three cells) as finite-sample variance on bounded exploratory
panels.

\section{Reconstructing the exploration and pilot decisions}
\label{sec:results}

\subsection{Calibration coverage and reasons for rejection}

Our secondary analysis script rebuilds all 76 configuration-by-seed loss
contrasts for the \RoneCells-cell exploration directly from archived
metrics before verifying the published cell and seed summaries
(\artifact{R1}). Pilot reconciliation follows a parallel path. Across 66
run-metrics files and 126 saved candidate-probe records, the script
recomputes paired utility and gain differences, null-adjusted property and
structure probe advantages, and composite eligibility flags against the
archived verification report (\artifact{P4}). All checks operate strictly
on saved numerical outputs, and Table~\ref{tab:cohorts} summarises the
resulting experimental units and failure counts.

\begin{table}[t]
\centering\small
\begin{tabularx}{\linewidth}{@{}lXX@{}}
\toprule
 & Exploration & Defence-redesign pilot \\
\midrule
Candidate configurations & \RoneCells & \PfourCells \\
Repetitions per candidate & 4 training seeds & 3 paired seeds \\
Recorded scope & Frozen exploratory rules & Frozen exploratory pilot \\
Qualified breach or failed composite & \RoneQualifiedBreaches & \PfourPrivacyFailures \\
Privacy undetermined & \RoneUndetermined & \PfourPrivacyUndetermined \\
Utility failures & \RoneUtilityFailures & \PfourUtilityFailures \\
Gain failures & \RoneGainFailures & 0 \\
Eligible candidate & None & None \\
\bottomrule
\end{tabularx}
\caption{Configuration counts and gate outcomes across both multi-cell
campaigns. Rejection causes in the \RoneCells-cell exploration overlap,
whereas the seven-cell pilot applies an all-bindings composite privacy
policy across its candidate mechanisms (\artifact{R1}, \artifact{P4}).}
\label{tab:cohorts}
\end{table}

Qualified privacy breaches eliminate \RoneQualifiedBreaches\ candidates
outright. Independent utility or remote-gain failures account for the
remaining \RoneUndetermined\ configurations and leave zero viable splits
under the frozen exploratory rules. Even if all 20 unqualified probes were
rebuilt and passed calibration, newly qualified measurements could only add
further breach flags without overturning any recorded rejection.

Uncertainty estimators also differ structurally between the two campaigns.
Exploratory cells apply two-sided 95\% Student-$t$ intervals over four
independent seeds, whereas pilot utility gates enforce a one-sided 95\%
upper Student-$t$ bound ($t_{0.95,2}=2.919986$) across three paired seeds.
Both conventions remain distinct throughout our reanalysis scripts.
Disjoint combinations of rejection causes appear separately in
Table~\ref{tab:intersections} (Appendix~\ref{app:checks}) so readers can
inspect exact failure overlaps without conflating heterogeneous confidence
rules.

\begin{figure}[t]
\centering
\includegraphics[width=\linewidth]{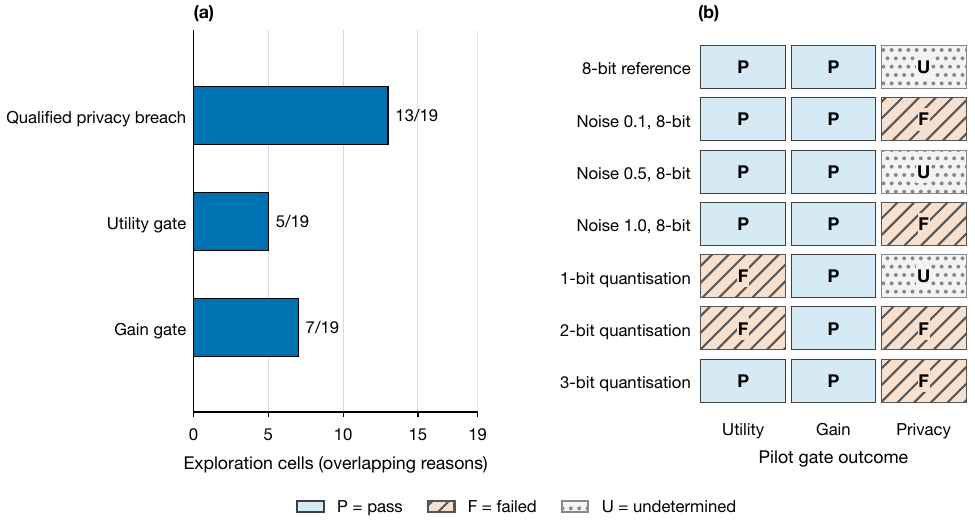}
\caption{Gate outcomes for the \RoneCells-cell exploration and seven-cell
defence pilot. Overlapping bars in panel~(a) tally qualified breaches,
utility losses, and gain deficits across four-seed exploratory candidates
(\artifact{R1}), while panel~(b) decomposes three-seed paired pilot cells
across utility, gain, and composite privacy rules (\artifact{P4}).}
\label{fig:campaigns}
\end{figure}

Separating atomic probe readouts from composite campaign labels clarifies
what happened inside the defence pilot. Atomic leakage stays visible. Every
pilot cell recorded calibrated next-byte structure breaches ($0.097$ to
$0.112$, with all 21 of 21 seed bindings qualified) and qualified
punctuation-property advantages ($0.79$ to $0.89$), yet when one or two
seed bindings for the punctuation-property probe stepped outside the
$\pm 0.02$ null band, the all-bindings rule labelled the composite cell
undetermined even though concrete calibrated leakage was already measured.
Figure~\ref{fig:campaigns} displays gate outcomes for both multi-cell
campaigns side by side without collapsing their distinct reporting rules.

\subsection{The narrow utility margin still passes}

\begin{figure}[t]
\centering
\includegraphics[width=\linewidth]{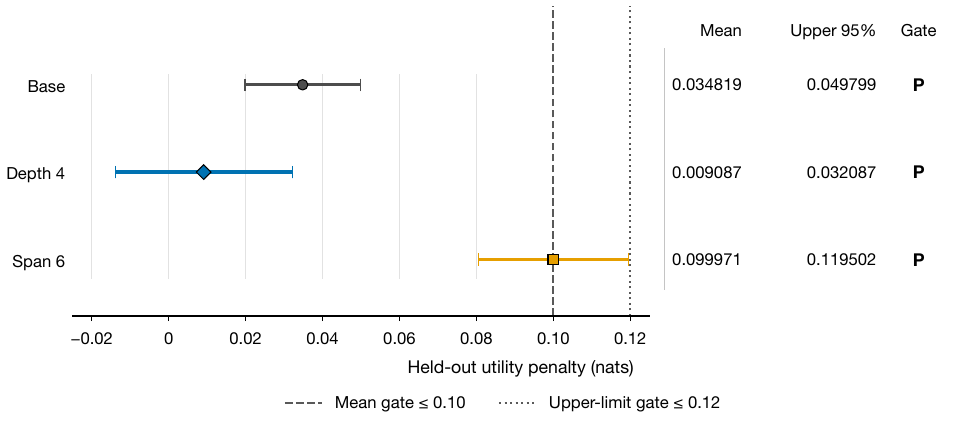}
\caption{Utility loss deltas and descriptive two-sided 95\% Student-$t$
intervals from Eq.~\eqref{eq:interval} across four training seeds on fixed
evaluation text (\artifact{R1}). Separate vertical lines mark the
$0.10$-nat mean cutoff and the $0.12$-nat upper-endpoint cutoff, which
explains why \texttt{span-6} clears both utility criteria even though its
upper interval crosses $0.10$.}
\label{fig:utility}
\end{figure}

Exploratory utility gating requires both a four-seed mean cross-entropy
increase of at most $0.10$ nats relative to baseline and an upper 95\%
Student-$t$ confidence bound of at most $0.12$ nats. Four training seeds
yield the descriptive interval
\begin{equation}
 \bar u\;\pm\;t_{0.975,3}\,\frac{s_u}{\sqrt{4}},
 \qquad t_{0.975,3}\approx3.182446.
 \label{eq:interval}
\end{equation}
Fixed evaluation panels define the underlying loss surface across all
configurations without multiplicity correction. Figure~\ref{fig:utility}
plots both acceptance cutoffs side by side.

Configuration \texttt{span-6} illustrates how dual-statistic rules behave
near a boundary (\artifact{R1}). Across its four seeds, \texttt{span-6}
attains a mean loss increase of $0.0999705344$ nats ($2.9466\times10^{-5}$
nats below the $0.10$ cap) and an upper confidence endpoint of
$0.1195024415$ nats ($4.9756\times10^{-4}$ nats below the $0.12$ limit).
Both utility requirements pass. Archived logs reject \texttt{span-6} solely
for failing remote-learning gain and triggering a qualified privacy breach,
while \texttt{depth-4} posts a smaller four-seed mean loss delta than the
base split within this exploratory cohort.

\section{Checking the scope of privacy claims}
\label{sec:scopechecks}

\subsection{Attack success under broader access}

Adversary privileges govern when empirical attack outcomes transfer across
threat models. A simple monotonic relationship separates demonstrated
breaches from empirical non-detections.

\begin{samepage}
\begin{proposition}[Transfer under nested observations]
\label{prop:nested}
Fix a target distribution, recovery metric, and computational budget.
Suppose observer $B$ receives every value visible to observer $A$ and can
apply every budget-admissible attack for $A$ after projecting onto $A$'s
observation space. There then exists a budget-admissible attack for $B$
whose expected recovery matches or exceeds that of any individual attack
available to $A$. Non-detection by a tested attack under $A$ places no
non-trivial upper bound on recovery under $B$.
\end{proposition}
\begin{proof}
Projecting $B$'s input onto $A$'s view and running $A$'s attack reproduces
the exact output distribution and recovery score within budget. Existence
of that projected attack leaves open whether a particular gradient
optimiser finds it, while failure of one tested attack under $A$ leaves
stronger uses of $B$'s extra state unconstrained.
\end{proof}
\end{samepage}

Shared-key orthogonal wrapping illustrates why
Proposition~\ref{prop:nested} invalidates careless threat-model upgrades
(\artifact{E8}). Multiplying hidden states by an orthogonal matrix $Q$
transforms boundary traffic via
\begin{equation}
 h'=hQ,\qquad h=h'Q^{\mathsf T}.
 \label{eq:rotation}
\end{equation}
Who holds $Q$ defines the security boundary. While a passive network tap
lacking $Q$ sees rotated coordinates $h'$, an authorised cloud server that
stores or derives $Q$ in host RAM to compute $h=h'Q^{\mathsf T}$ breaks the
no-key premise immediately if compromised. Archived experiment accounts and
collection manifests restrict the restoration result to passive wire and
capture observers (\artifact{E8}, \artifact{ER}). Recent attacks that
invert statistical and permutation obfuscation in third-party LLM serving
highlight the same host-exposure boundary \cite{permattack2025}.

Orthogonal maps also preserve Euclidean norms and token inner products
exactly. Coordinate mismatch explains why a position-wise MLP decoder fails
on $h'$ while relational Gram geometry stays intact across the sequence.
Stronger known-plaintext alignment solvers inside the robustness archive
exploit labelled pairs to recover the subspace at depth-dependent sample
sizes (\artifact{E8}), consistent with independent findings that deeper
transformer cuts retain recoverable internal states
\cite{depth2025,actinv2026,bettisplit2026}.

\begin{table}[t]
\centering\small
\begin{tabular}{@{}rrrr@{}}
\toprule
Depth & Estimated $K_{50}$ & Measured bracket & $K_{50}/5120$ \\
\midrule
1 & 258.32 & 200 to 300 & 0.05045 \\
4 & 355.81 & 300 to 400 & 0.06949 \\
8 & 617.93 & 600 to 700 & 0.12069 \\
\bottomrule
\end{tabular}
\caption{Interpolated absolute 50\% token-recovery crossings ($K_{50}$)
from the 27B ($H=5120$) robustness artifact across split depths
(\artifact{E8}). Normalised ratios $K_{50}/5120$ span a $2.39$-fold range
across depth without confidence intervals on the interpolated crossings.}
\label{tab:k50}
\end{table}

Aggregating sessions can obscure peak exposure as well (\artifact{E8}).
Under per-session rotation at depth~1, the pooled top-1 recovery score is
$3.71\%$, yet constituent sessions reach $5.73\%$, $3.65\%$, and $2.34\%$.
Pooled averages understate the leakiest session. Because known-plaintext
solvers extract sensitive partial spans well below the absolute 50\%
crossing in Table~\ref{tab:k50}, interpolated $K_{50}$ brackets track
empirical attack scaling in the lab without defining safe operational key
lifetimes.

\subsection{Comparing rotation cadences}

Archived deployment logs separate a three-session Mistral-7B per-session
rotation test from a later eight-run Mistral-NeMo-12B per-epoch ratcheting
matrix (\artifact{ER}). Per-session swaps add negligible latency. Within
the 12B per-epoch matrix, greedy-decode identity holds in 20 of 24 sessions
and four of the eight runs carry documented admissibility caveats, while
the \texttt{N=256} arm executes zero rotations across its 60-token decode
horizon and acts as a static control (though its zero-RTT throughput still
drops from $7.6$--$7.7$ to $6.3$--$6.4$ tokens/s due to harness overhead).
Wire decoders scored zero top-1 recovery on rotated epochs within that
bounded capture setup.

Frequent ratcheting imposes steep latency penalties (\artifact{ER}). At
cadence \texttt{N=16} with zero emulated RTT in session~1, recorded
throughput drops from $7.6$ tokens/s on the static baseline to $2.8$
tokens/s under ratcheting (a $63.16\%$ reduction at saved precision),
whereas the evidence-triggered budget arm runs near $5.5$ to $5.6$ tokens/s
at zero emulated RTT (and $2.6$ to $2.9$ tokens/s at 80\,ms RTT). Model
scale, rotation cadence, and RTT twins remain distinct experimental factors
across those extracted logs.

Provenance reconciliation explains why two internal documents disagreed
about this matrix (\artifact{ER}). An early summary flagged the entire
ratchet sweep invalid before a subsequent RTT-twin review readmitted the
supporting runs alongside explicit identity, throughput, and cloud-RAM key
caveats. Following our precedence rule, we adopt the later collection
manifest together with every attached operational restriction.

\subsection{Reading mutual information estimates}

Companion boundary evaluations report near-zero Mutual Information Neural
Estimation (MINE) \cite{belghazi2018} readouts on defended latent splits
(\artifact{MI}). For any measurable critic $T$ with finite expectations
($\mathbb E_{P_{XZ}}[|T|]<\infty$ and $\mathbb E_{P_XP_Z}[e^{T}]<\infty$),
the population Donsker--Varadhan variational objective obeys
\begin{equation}
 L_T=\mathbb E_{P_{XZ}}[T(X,Z)]-
 \log\mathbb E_{P_XP_Z}[e^{T(X,Z)}]\le \operatorname{MI}(X,Z),
 \label{eq:dv}
\end{equation}
where $\log$ is the natural logarithm and $\operatorname{MI}(X,Z)$ denotes
true mutual information in nats. That inequality bounds leakage strictly
from below. In addition to severe finite-sample bias and exponential
variance in variational estimators \cite{mcallester2020,song2020},
Eq.~\eqref{eq:dv} points in the wrong direction to certify an upper leakage
budget, just as weak empirical attacks in differential-privacy auditing
never prove small $\varepsilon$ \cite{jagielski2020}.

A trivial analytic channel exposes the flaw immediately. Let input $X$ be
uniform over $m$ discrete symbols, transmit plaintext $Z=X$, and evaluate
constant critic $T=0$. Critic output $L_T$ equals zero identically even
though $\operatorname{MI}(X,Z)=\log m$ nats ($8$ full bits at $m=256$).
Pairing the MINE readout with a second lower-bound estimator (such as the
companion study's calibrated cross-entropy probe bound) leaves the
inequality direction unchanged, and the companion draft's own limitations
section acknowledges that neither estimator furnishes an upper bound
(\artifact{MI}).

We therefore treat near-zero critic readouts as finite non-detections by a
particular neural family while leaving true channel leakage open in both
directions. Architectural ablations require identical restraint
(\artifact{MI}). Observing comparable task metrics across a handful of
invariant MLP and radial remote modules demonstrates empirical saturation
within the tested training budget, whereas proving a universal
information-theoretic capacity ceiling would require bounding every
admissible remote function family.

\section{Relationship to existing evidence and methods}
\label{sec:related}

Split learning was originally introduced to train collaborative models
without exporting raw client records \cite{vepakomma2018}, and recent
surveys catalogue a rapidly growing literature adapting split architectures
to LLM fine-tuning \cite{splitsurvey2026}. Backward tensors still leak
text. Modern reconstruction attacks recover private inputs from exchanged
gradients and activations \cite{gisurvey2022,dager2024,fedspy2026}. Within
our examined programme, adding returned gradients in a controlled GPT-2
benchmark lifts exact-document recovery from $13.71\%$ to $37.77\%$ while
increasing token recovery by $3.17\pp$ \cite{gradients2026}, and
client-returned loss gradients nullify synthetic decoys through exact-zero
support masks \cite{paper1}. Temporal gradient correlations leak
trajectories through similar structural channels in reinforcement learning
\cite{trace2026}. Here we retain each cited measurement under its original
attribution and denominator.

Methodological overlap with LLM-Attack Suite calls for clear boundaries
\cite{suite2026}. That framework already specifies observation contracts,
undefended controls, saved predictions, missing-attack disclosures, and
harness checks. Our work operates one layer higher. By reconciling
cross-campaign decision tables where qualified breaches, uncalibrated
probes, shifted labels, and host-key assumptions collide, we isolate which
programme-level conclusions survive without claiming diagnostic benchmarks
on unseen external repositories.

External studies warn repeatedly that weak evaluators manufacture false
privacy. Wu et al.\ \cite{adaptive2022} break federated defences calibrated
against static attacks, while Smith et al.\ \cite{mia2026} measure
$1.16$-fold to $6.59$-fold higher leakage on FaceScrub under adaptive
attack optimisation (and separately show that changing the external
evaluator's feature basis alters measured leakage). Threat-model and budget
sensitivity appears across multimodal and LLM benchmarks as well
\cite{labelbudget2026,ablation2026}. Layer-wise analyses independently
confirm that deeper transformer splits retain recoverable internal states
\cite{depth2025,actinv2026,bettisplit2026}. Evaluator scepticism still cuts
only one way, because disqualifying 20 uncalibrated probes in our
\RoneCells-cell exploration leaves all \RoneQualifiedBreaches\ breaches
caught by the four calibrated probes intact.

Recent defence studies reinforce both our host-custody and
variational-bound findings. Jin et al.\ \cite{privpair2026} show that
perturbation alone struggles during federated split-LLM fine-tuning and
propose client obfuscate-and-recover adapters, Chen et al.\
\cite{mixguard2026} introduce mixup-based activation obfuscation for split
LLMs, and Zhang et al.\ \cite{aegis2026} investigate generative gradient
masking in medical federated learning. Key custody remains decisive. Thomas
et al.\ \cite{permattack2025} invert permutation-based third-party LLM
inference schemes and illustrate why invertible coordinate transforms
demand explicit host-memory accounting. On the information-theoretic side,
Noorbakhsh et al.\ \cite{mitech2026} apply mutual-information
regularisation during collaborative split inference and Gu et al.\
\cite{pidi2026} incorporate a mutual-information regulariser during
split-LLM fine-tuning, while formal statistical limits on neural
mutual-information estimators \cite{belghazi2018,mcallester2020,song2020},
paired with empirical differential-privacy auditing \cite{jagielski2020},
explain why near-zero critic scores never certify small upper leakage
budgets.

\section{Limitations and implications for further evaluation}
\label{sec:limits}

\paragraph{Selection and dependence.}
Case selection occurred retrospectively after all campaign outcomes had
already been observed. Shared codebases, corpora, checkpoints, and design
decisions link every episode inside one group. Archive tallies describe
this collection alone, and measuring false-positive or false-negative
classification rates on external papers would require a prospectively
registered validation cohort.

\paragraph{Evidence available for recomputation.}
Executable scripts recompute seed contrasts, confidence bounds, and
composite gate decisions wherever raw CSV or JSON arrays exist in the
package. Some upstream artifacts remain source-reported. Decoy-gradient
frame counts and portions of the boundary-replacement review come from
their primary manuscripts and archived review verdicts without refitting
attackers or rescoring tensor captures. File-level SHA-256 digests verify
byte identity across all \PinnedInputs\ consumed records, and untraced
constants lacking a primary calibration curve stay excluded.

\paragraph{Qualification and uncertainty.}
Probe qualification binds strictly to one declared target, channel, defence
mechanism, and budget. Seed cohorts remain small. Four-seed two-sided
Student-$t$ intervals in the exploration and three-seed one-sided upper
bounds in the defence pilot condition on fixed evaluation panels without
multiplicity adjustments, while the unexecuted eight-seed power figure
serves purely as a variance-sensitive planning bound. Positive-control
canary floors, permutation-null bands, acceptance gates, and minimum
detectable leak rates measure distinct properties even when two constants
share the value $0.02$.

\paragraph{Security and utility.}
Empirical non-detection reflects a bounded attacker and compute budget. Low
pooled passive recovery, invertible wrappers whose keys sit in cloud RAM,
and near-zero variational lower bounds leave compromised-host secrecy
unproved. Utility gates cover only their scored endpoints, since
short-context cross-entropy can pass while teacher-forced answer-token
accuracy drops sharply, and evaluating noise or quantisation fairly
requires sweeping perturbation strength against explicit task-quality caps.

Explicit threat contracts enable testable future benchmarks. Deliberately
injected channel omissions and label shifts can run alongside genuinely
leaky defences under frozen compute budgets. Prospective validation trials
on held-out model families and corpora can then quantify how reliably
automated admissibility checks catch false privacy assurance while
preserving confirmed breaches.

\section{Conclusion}

Re-examining six linked investigations from one split-LLM programme
separates broken measurements from broken defences. Two early assurances
failed on method. Omitted client-returned gradients and off-by-one token
labels undermined those initial claims, yet \RoneQualifiedBreaches\
qualified breaches in the \RoneCells-cell exploration alongside $40\times$
punctuation-property excesses and $5\times$ next-byte structure excesses in
the seven-cell pilot stand as sound negative evidence against the tested
defences despite incomplete probe qualification elsewhere. Scope checks
likewise confine shared-key orthogonal wrappers to passive network links,
document the $63.16\%$ throughput penalty of rapid per-epoch ratcheting,
and show why variational mutual-information lower bounds near zero leave
true leakage unconstrained above. Tying every privacy verdict to its
observed channels, key custody, calibrated controls, and seed-level
uncertainty establishes exactly what archived split-training experiments
support.

{\emergencystretch=2em
\paragraph{Data and code availability.}
The manuscript source, reanalysis scripts, derived tables, and archived
evidence reside in the companion repository at \repourl{}
\cite{artifacts2026}. Python recomputation scripts live under
\path{analysis/}, derived tables and SHA-256 input manifests under
\path{data/}, LaTeX sources and vector figure generators under
\path{paper/}, and primary records under \path{evidence/} and
\path{reproduction/}. Appendix~\ref{app:artifacts} maps each retained
statistic to its exact repository-relative path. Every consistency check
runs offline on the Python standard library without GPU retraining.\par}

\appendix
\section{Sources and reproducibility}
\label{app:artifacts}

{\emergencystretch=3em
Relative paths below resolve from the repository root (\repourl{}). Byte
counts and SHA-256 digests for all \PinnedInputs\ pinned inputs appear in
\path{data/source_manifest.json}, paired with a flat lookup table in
\path{data/source-index.csv}. Every path cited below is directly hashed in
that manifest while preserving the original campaign manifests as separate
provenance records.

\begin{description}
\item[\hypertarget{artifact-E1}{Decoy-gradient experiment.}]
Primary text and tables reside in
\path{evidence/decoy-gradient/paper1-main.tex} and
\path{evidence/decoy-gradient/E1_GRADIENT_CHANNEL_2026-08-22.md}, matching
public preprint \cite{paper1} on the 4,096-frame support leak, the $+0.65$
to $+1.50\pp$ token-probe gain, the 1,024-frame per-run mitigation runs,
and the $0.01$-nat clipping-and-noise utility penalty. Frame tallies remain
source-reported here without re-parsing raw training logs.

\item[\hypertarget{artifact-B1}{Boundary-replacement review.}]
Archived findings live in
\path{evidence/boundary-replacement/independent_evidence_review_2026-09-09/VERDICTS.md}
and its companion recomputed JSON
\path{evidence/boundary-replacement/independent_evidence_review_2026-09-09/recomputed_evidence.json}.
Those two review records trace the shifted sequence alignment
$(x_1,\ldots,x_{T-1})$ versus $(x_2,\ldots,x_T)$, the six zero-return
baseline comparisons, and the $16.64\pp$ answer-token drop to their
original evaluation outputs. Our manifest hashes both review files
directly, while \path{recomputed_evidence.json} records the SHA-256 digests
of the 66 upstream evaluation artifacts.

\item[\hypertarget{artifact-R1}{Privacy, utility and gain exploration.}]
Raw exploratory outputs come from
\path{reproduction/trilemma-campaign/evidence/r1-exploration-20260917/analysis/exploratory_analysis.json},
\path{reproduction/trilemma-campaign/evidence/r1-qualification-20260916/attack-spec-frozen.json},
\path{reproduction/trilemma-campaign/evidence/r1-qualification-20260916/calibration_analysis_summary.json},
\path{evidence/trilemma-campaign/r1_cell_results.csv}, and
\path{evidence/trilemma-campaign/r1_seed_results.csv}. Our script rebuilds
all 76 configuration-by-seed contrasts, Student-$t$ endpoints, probe
qualifications, and gate outcomes, while reading the unexecuted planning
bound from
\path{reproduction/trilemma-campaign/evidence/r1-exploration-20260917/confirmation_power.json}
and \path{evidence/trilemma-campaign/R1_CONFIRMATION_REVIEW.md}.

\item[\hypertarget{artifact-P4}{Defence-redesign pilot.}]
Pilot records comprise
\path{reproduction/defense-redesign-pilot/outputs/runs/pilot-20260922-a/verification.json},
the adjacent generated report and 66 run-metrics files, and
\path{evidence/defense-redesign-pilot/PILOT_DECISION.md}. From those saved
candidate and null arrays, our script recomputes three-seed paired utility
bounds, remote-gain contrasts, punctuation-property and next-byte structure
advantages, and all-bindings composite verdicts.

\item[\hypertarget{artifact-E8}{Orthogonal-wrapper experiment.}]
Wrapper measurements come from
\path{reproduction/structural-e8/paper-data/collected/h100-2/issues/issue_e8_27b_20260807/e8_robustness_27b_bf16_densek_20260807T230529Z.json}
and \path{evidence/structural-e8/EXPERIMENT_STORY.md}. Extracted fields
cover undefended baselines, depth-specific $K_{50}$ interpolation brackets,
pooled $3.71\%$ recovery, and individual session scores up to $5.73\%$.

\item[\hypertarget{artifact-ER}{Rotation deployment study.}]
Deployment records draw from
\path{reproduction/structural-e8/paper-data/collected/toula/er_20260811/COLLECTION_MANIFEST.json},
accompanying session-identity JSONs, and
\path{evidence/structural-e8/EVIDENCE_EXECUTION_LEDGER.md}. Across eight
runs and 24 sessions, session~1 at \texttt{N=16} and zero emulated RTT logs
$2.8$ versus $7.6$ tokens/s and establishes the $63.16\%$ throughput drop
at recorded precision.

\item[\hypertarget{artifact-MI}{MI-estimator analysis.}]
Variational critic records reside in
\path{evidence/privacy-contract/privacy-contract-draft.tex} and
\path{evidence/privacy-contract/KEY_NUMBERS.md}, paired with finite
architecture sweeps in
\path{evidence/structural-e8/LATENT_CAMPAIGN_REPORT_V7_V9.md}. Near-zero
critic scores are cited as source-reported alongside Eq.~\eqref{eq:dv} and
its accompanying constant-critic counterexample.
\end{description}
}

\section{Checking the derived records}
\label{app:checks}

Running the standalone Python script from the repository root regenerates
every derived table, macro file, and consistency check:
\begin{quote}\small\ttfamily
python3 analysis/recompute.py
\end{quote}
Execution relies exclusively on the Python standard library and writes
derived outputs under \path{data/} (syncing count macros to
\path{paper/results_macros.tex}) without touching model checkpoints, tensor
captures, or cryptographic keys. Saved logs verify \ConsistencyChecks\
checks across the \PinnedInputs\ hashed inputs, comparing recomputed seed
contrasts, confidence limits, and gate flags against the archived
campaign tables. Table~\ref{tab:intersections} reports the resulting
disjoint rejection-reason intersections for the \RoneCells-cell
exploration.

\begin{table}[htb]
\centering\small
\begin{tabular}{@{}lr@{}}
\toprule
Rejection reason combination & Configurations \\
\midrule
Qualified privacy breach only & 10 \\
Gain failure and qualified privacy breach & 3 \\
Utility and gain failure & 3 \\
Utility failure only & 2 \\
Gain failure only & 1 \\
\midrule
Total & \RoneCells \\
\bottomrule
\end{tabular}
\caption{Disjoint rejection-reason intersections across all
\RoneCells\ exploratory configurations (\artifact{R1}). Partitioning the
marginal bars in Figure~\ref{fig:campaigns}(a) confirms that all
\RoneUndetermined\ configurations without a qualified privacy breach fail
utility or learning-gain gates independently, as stored in
\texttt{data/reanalysis\_results.json}.}
\label{tab:intersections}
\end{table}

Derived JSON outputs preserve per-campaign confidence conventions, scoring
denominators, and operational caveats. Before writing vector PDFs,
\path{paper/figs/make_figures.py} cross-checks \path{data/r1_cells.csv} and
\path{data/p4_cells.csv} against the archived exploration cell and seed
CSVs, the generated pilot report
\path{reproduction/defense-redesign-pilot/outputs/runs/pilot-20260922-a/report.md},
and the seven pilot run manifests. The full build recipe
\path{paper/build.py} verifies all \PinnedInputs\ input hashes, runs
\path{analysis/recompute.py}, compiles the manuscript with pdfLaTeX and
BibTeX, and packages a self-contained \path{paper/arxiv-source.zip} bundle.

\FloatBarrier
\bibliographystyle{plain}
\bibliography{refs}
\end{document}